\documentclass[,preprint,aps,prd]{revtex4}
\usepackage{epsfig}
\usepackage{epsf}
\usepackage{amsmath}
\usepackage{amssymb}
\usepackage{graphicx}

\begin{document}



\title{Entanglement of Uniformly Accelerating Schr\"{o}dinger, Dirac, and Scalar Particles}

\author{W. L. Ku}
\author{M.-C. Chu}
\affiliation{   Department of Physics and Institute of Theoretical Physics, the Chinese University of Hong Kong, Hong Kong SAR, PRC}

\date{\today}

\begin{abstract}
 We study how the entanglement of an entangled pair of particles is
 affected when one or both of the pair is uniformly accelerated, while the
 detector remains in an inertial frame. We find that the
 entanglement is unchanged if all degrees of freedom are considered.
 However, particle pairs are produced when a relativistic particle is
 accelerated, and more bipartite systems emerge, the entanglements of some of
 which may change as the acceleration. In particular, the entanglement of a
 pair of accelerating fermions is transferred preferentially to the produced
 antiparticles when the acceleration is large, and the entanglement transfer
 is complete when the acceleration approaches infinity. However, no such
 entanglement transfer to the antiparticles is observed for scalar particles.
\end{abstract}


\maketitle

\section{Introduction}
\label{sect:Intro}
  Entanglement is an important property of quantum mechanical systems.
  It is useful in the field of quantum information and quantum computing,
  such as in quantum teleportation~\cite{qutele}. It also
  finds many applications in quantum control~\cite{qc} and quantum
  simulations~\cite{qs}. Studying quantum entanglement in relativistic systems may give us
  insights on the relationship between quantum mechanics and general
  relativity. It has been shown that entanglement is
  Lorentz invariant~\cite{adami,alsing}. However, an accelerating observer
  measures less entanglement than an inertial observer in both the
  scalar~\cite{tele,nb} and fermion~\cite{nf} cases. This degradation in
  entanglement is due to the splitting of space-time, as a
  result of which the vacuum observed in one frame can become excited in
  another frame - the case of Unruh
  effect~\cite{unruh}. Classically, the trajectory of a uniformly
  accelerating particle observed by an inertial observer is the
  same as that of an inertial particle measured by a uniformly accelerating
  observer with appropriate acceleration. We are interested in how the acceleration of particles affects the
  entanglement of the originally entangled states.

  In this paper, we analyze the entanglement of accelerating
  particles in three cases: a non-relativistic wave packet, scalar and fermion
  particles. We compare the entanglement of accelerating
  particles, as seen by an inertial detector, with that of inertial particles
  observed by an accelerating detector~\cite{tele,nb,nf,zhang}. We find that when all
  degrees of freedom are considered, the entanglement is unchanged. However, pair
  production occurs when a relativistic particle accelerates, and there are new
  bipartite systems. We find that the entanglements of some of the new bipartite
  systems can depend on the acceleration. In particular, a pair of accelerating
  fermions transfer their entanglement preferentially to the produced antiparticles
  when the acceleration is large, and the entanglement transfer is complete when the
  acceleration approaches infinity. However, no such entanglement transfer to the
  antiparticles is observed for scalar particles.

  This paper is organized as follows. In Sec.~\ref{sec2}, we
  add a potential term in the Schr\"{o}dinger equation that would lead to
  uniform acceleration in the classical limit. Then we construct a
  wave packet solution and a two-body entangled wave function, and
  we calculate the entanglement by Schmidt decomposition~\cite{sd} as a
  function of acceleration. In Sec.~\ref{formal}, we add the same
  potential term in the Klein-Gordon equation and Dirac equation. A wave packet solution
  can be obtained~\cite{kgwp_0} to give an intuitive picture of how a relativistic particle
  accelerates and pair production occurs. We quantize the fields
  and use the in/out formalism to calculate the pair production,
  and we introduce the logarithmic negativity~\cite{lne} to calculate the
  entanglements in different bipartite systems. In Sec.~\ref{sec4}, we consider the case
  when one or both Dirac particles are accelerated. Entanglements between different
  degrees of freedom are calculated and entanglement transfer to the antiparticles will be
  shown. In Sec.~\ref{sec3}, we calculate the particle spectrum for scalar
  particles, and the results are compared with those observed by an
  accelerating observer. We also repeat the calculation of entanglements in Sec.~\ref{sec4}
  but for scalar particles.
  A summary and discussion of results and further work is given in Sec.~\ref{sec5}.

\section{Accelerating schr\"{o}dinger particles} \label{sec2}

A free non-relativistic particle with mass $m$ represented by a
gaussian wave packet

\begin{eqnarray}
\psi_{0}(x,t)=\frac{(8b/\pi)^{1/4}}{\sqrt{4b+2i
t/m}}\exp\left(-\frac{x^{2}}{4b+2i t/m}\right)
\end{eqnarray}
satisfies the Schr\"{o}dinger equation~$(\hbar=c=1)$,

\begin{eqnarray}
i\frac{\partial \psi}{\partial
t}=\frac{-1}{2m}\frac{\partial^{2}\psi}{\partial x^{2}}.
\end{eqnarray}
It is reasonable to assume that the center of the wave packet
follows the classical trajectory of the corresponding
particle~\cite{accwp}. Therefore, when a linear potential is added
to the Schr\"{o}dinger equation,

\begin{eqnarray}
i\frac{\partial \psi}{\partial
t}=\frac{-1}{2m}\frac{\partial^{2}\psi}{\partial x^{2}}-max\psi,
 \label{acceqn}
\end{eqnarray}
we use an ansatz of the form
\begin{eqnarray}
\psi(x,t)=\psi_{0}(x-x_{0}-v_{0}t-\frac{1}{2}at^{2},t)e^{iS(x,t)},
\end{eqnarray}
which when substituted into Eq.~(\ref{acceqn}) produces an
accelerating wave packet with

\begin{eqnarray}
\frac{1}{m}S(x,t;v_{0},a)=v_{0}x+axt-\frac{1}{2}av_{0}t^{2}-\frac{1}{6}a^{2}t^{3}-\frac{1}{2}v_{0}^{2}t,
\end{eqnarray}
where $ v_{0}$ and $a$ are treated as parameters that represent the
initial velocity and acceleration in the classical limit.

\begin{figure}
\includegraphics{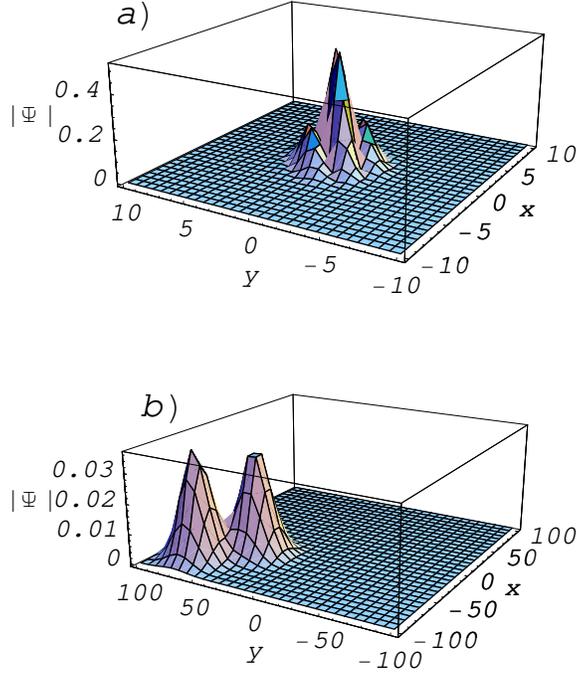}
\caption{\label{fig:accwp_0}  a) A 3D plot of the absolute value of
the two-body wave function $\Psi_{+}$ Eq.~(\ref{2bdwf}) at $t$ = 0,
for $v_{1}$ = -1, $v_{2}$ = 1, $x_{1}$ = $x_{2}$ = 0, $m_{1}$ =
$m_{2}$ = 1, $a_{1}$ = -0.5, $a_{2}$ = 0.5. b) Same as
Fig.~\ref{fig:accwp_0}a, but for $t$ = 15.}
\end{figure}

\begin{figure}
\includegraphics{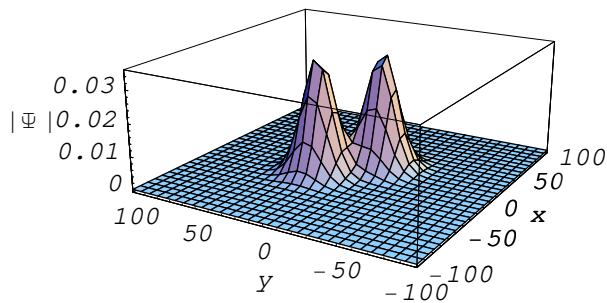}
\caption{\label{fig:accwp_2} Same as Fig.~\ref{fig:accwp_0}b, but
 for $a_{1}$ = $a_{2}$ = 0 (free evolution).}
\end{figure}

We write down the two-body entangled wave functions as follows,

\begin{eqnarray}
\label{2bdwf} \Psi_{\pm}(x,y,t)&=&
N[\psi(x,t;v_{1},a_{1})\psi(y,t;v_{2},a_{2})\pm
\psi(x,t;v_{2},a_{1})\psi(y,t;v_{1},a_{2})],
\end{eqnarray}
where $N$ is the normalization factor. For simplicity, we have just
set the masses, initial positions and widths of the two particles in
the wave packet to be the same. If $a_{1}$ or $a_{2}$ $\neq$ 0, the
two-body wave function accelerates in a specific direction~(see
Figs.~\ref{fig:accwp_0} and \ref{fig:accwp_2} for examples).

\begin{figure}
\includegraphics{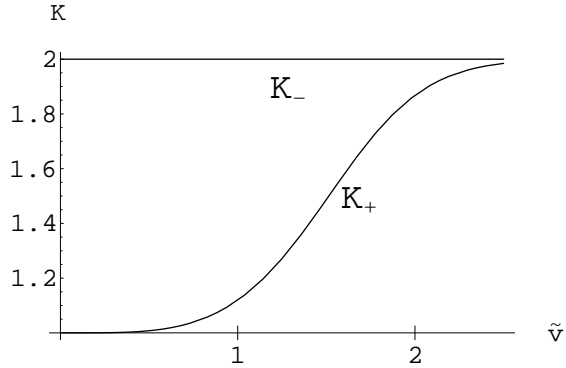}
\caption{\label{fig:schno}The Schmidt numbers $K_{\pm}$ as function
of the relative velocity of the two particles, made dimensionless by
~$\tilde{v}\equiv vm\sqrt{b}$, of the two-body wave functions
$\Psi_{\pm}$.}
\end{figure}

We can calculate the purity $P$ of the wave function

\begin{eqnarray}
P&=&\int\int\int\int \Psi(x,y,t)\Psi(x',y',t)
\Psi^{*}(x,y',t)\Psi^{*}(x',y,t)dxdx'dydy',
\end{eqnarray}
and hence the Schmidt number: $K\equiv 1/P$. For a product state,
$K=1$, and for entangled states, $K$ is greater than 1. The
entanglement depends on the relative velocity of the two particles
$v$ instead of the velocity of each of the particles. Therefore, we
can choose to calculate the entanglement in the frame that $v_{1}$ =
0 and $v_{2}$ = $v$. The Schmidt number for $\Psi_{+}(x,y,t)$ is
calculated to be~(see Fig.~\ref{fig:schno})

\begin{eqnarray}
K_{+}=\frac{2}{1+{4f}/{(1+f)^{2}}} ,
\end{eqnarray}
where $f=\exp(-\tilde{v}^{2})$ and $\tilde{v}\equiv v m\sqrt{b}$. In
this case, the entanglement depends only on the product
$bv^{2}m^{2}$ and is independent of the acceleration. This result
can be easily understood. In the wave function, $a$ appears always
as a product with $t$. As the Hamiltonian is Hermitian, the
evolution operator is unitary, and the entanglement is unchanged
under a unitary transformation. Thus we can choose to calculate the
entanglement at time $t$ = 0, and all the acceleration terms will
disappear. For $\Psi_{+}$, $K_{+}$ is equal to 1 when $v$ is equal
to 0, and $K_{+}$ is equal to 2 when $v$ tends to infinity. For
$\Psi_{-}$, $K_{-}$ is always equal to 2. Although the entanglement
is independent of acceleration in both cases, there is a difference
between the entanglements of $\Psi_{+}$ and $\Psi_{-}$. The
entanglement depends on the orthogonality of the two terms in
$\Psi_{\pm}$. For $\Psi_{+}$, a greater $v$ increases the
orthogonality of the two terms in the wavefunction. The minus sign
in $\Psi_{-}$ cancels the overlap region between the two terms in
Eq.~(\ref{2bdwf}), and the remaining parts are orthogonal. As a
result, the entanglement of $\Psi_{-}$ is always maximum in the
system.

\section{Relativistic Formalism}\label{formal}
\subsection{Quantization of Fields}
In order to accelerate a relativistic particle, the Klein-Gordon
(KG) or Dirac equation with an electric field is considered. A
strong electric field makes the vacuum unstable and leads to pair
production ~\cite{fs,wh,jsch}, which has been studied in the time
dependent gauge~\cite{pptg_1,pptg_2}, in Rindler coordinates
~\cite{ppr} and in a finite region ~\cite{ppf_1,ppf_2,ppf_3}.

The pair production of scalar particles can be understood in the
picture of a wave packet ~\cite{kgwp_0,kgwp_1}. A wave packet
uniformly accelerates from the far past, and then tunneling occurs
in the region when it meets the potential barrier. The transmission
wave packet represents the antiparticles while the reflected wave
packet represents the particles. Therefore, pair production occurs
in the tunneling region, and this solves the Klein paradox
~\cite{klein}. Although the wave packet formalism is more intuitive,
it is not clear how to construct a two-body entangled probability
density in the Klein-Gordon field. Thus, we quantize the field and
calculate the Bogoliubov coefficients in the in/out formalism.
\subsubsection{Scalar Particles}

The Klein-Gordon equation ~\cite{kgwp_0,kgwp_1} for a unit-charged
particle with mass $m$ in a uniform electric field $E$ is

\begin{eqnarray}
(D_{\mu}D^{\mu}+m^{2})\phi =0,\label{okg}
\end{eqnarray}
where $D_{\mu}=\partial_{\mu}+iA_{\mu}$ and the gauge is chosen to
be $A_{0}$ = $-Ex$ and $A_{x}$ = 0.  We assume
$\phi_{\omega}(t,x)=C\exp{(i\omega t)}\chi_{\omega}(x)$, and the
spatial part solutions are parabolic cylinder functions,
$D_{-a-\frac{1}{2}}(x)$. Details of obtaining the solutions are
shown in Appendix \ref{appa}. Then we classify the solutions in the
in/out basis \cite{kgwp_0,greiner}, such that we have two complete
bases to quantize the field as

\begin{eqnarray}
&&
\phi=\sum_{\omega}(a_{\omega}^{in}\phi_{\omega,p}^{in}+b_{\omega}^{in\dagger}\phi_{\omega,a}^{in*}),
\end{eqnarray}
or
\begin{eqnarray}
\phi=\sum_{\omega}(a_{\omega}^{out}\phi_{\omega,p}^{out}+b_{\omega}^{out\dagger}\phi_{\omega,a}^{out*}),
\end{eqnarray}
where the subscripts $p$ and $a$ label the particles and
antiparticles respectively. The operators
$a_{\omega}^{in}$($b_{\omega}^{in\dagger}$) and
$a_{\omega}^{out}$($b_{\omega}^{out\dagger}$) are the
annihilation~(creation) operators in the in-basis and out-basis, and
they are related by the Bogoliubov transformation,

\begin{eqnarray}
&& a_{\omega}^{out}=\alpha^{*}a^{in}_{\omega}-\beta^{*}
b_{\omega}^{in\dagger},
\\ && \nonumber
b_{\omega}^{out}=\alpha^{*}b_{\omega}^{in}-\beta^{*}
a_{\omega}^{in\dagger},
\end{eqnarray}
where

\begin{eqnarray}
&&
\alpha=\frac{\sqrt{2\pi}e^{-i\pi/4}e^{-\pi\mu^{2}/2}}{\Gamma(1/2+i\mu^{2})},
\\ && \nonumber
\beta=e^{i\pi/2}e^{-\pi\mu^{2}},
\end{eqnarray}
with $\mu^{2}=m^{2}/2E$, and

\begin{eqnarray}
|\alpha|^{2}-|\beta|^{2}=1. \label{scarela}
\end{eqnarray}

We can express the in-vacuum state as the linear combination of out
states \cite{kgwp_0} as,
\begin{eqnarray}
|0\rangle_{in}=\prod_{\omega}\frac{1}{\alpha}\exp\left[\left(-\frac{\beta^{*}}{\alpha}\right)a^{out\dagger}_{\omega}b^{out\dagger}_{\omega}\right]|0\rangle_{out}.
\end{eqnarray}
We let $\alpha=e^{i\phi_{1}}\cosh{r}$  and
$\beta=e^{i\phi_{2}}\sinh{r}$, where $0<r\lesssim0.88$ is a
parameter related to acceleration,  and we neglect the phase factors
which do not affect the following calculations of entanglement.
Taking the single-mode approximation, we get the in-vacuum state in
terms of the out states,

\begin{eqnarray}
&&
|0_{p}\rangle_{in}=\frac{1}{\cosh{r}}\sum_{n=0}^{\infty}\tanh^{n}{r}|n_{p}\rangle_{out}|n_{a}\rangle_{out}.
\end{eqnarray}
Similarly the one-particle state is
\begin{eqnarray}
|1_{p}\rangle_{in}
=\frac{1}{\cosh^{2}{r}}\sum^{\infty}_{n=0}\tanh^{n}{r}\sqrt{n+1}|(n+1)_{p}\rangle_{out}|n_{a}\rangle_{out}.
\end{eqnarray}

\subsubsection{Fermions}

For a unit-charged fermion with mass $m$ coupled to an uniform
electric field,
\begin{eqnarray}
[\gamma^{\mu}(p_{\mu}-A_{\mu})-m]\psi=0 \label{diraceqt},
\end{eqnarray}
where $A_{\mu}$ is the vector potential and $\gamma_{\mu}$ is the
gamma matrix. Eq.~(\ref{diraceqt}) can be reduced to two
Klein-Gordon equations which are shown in Appendix \ref{appb}. The
solutions are still the parabolic cylinder functions. The in/out
basis solution of the second order ODE is still the in/out basis
solution of the Dirac equation Eq.~(\ref{diraceqt}). Therefore, we
obtain the Bogoliubov coefficients, which have been calculated in
Ref.~\cite{niki},

\begin{eqnarray}
&& a_{n}^{out}=\alpha_{f} a_{n}^{in}-\beta_{f}^{*}b_{n}^{in\dagger},
\\ && \nonumber
b_{n}^{out\dagger}=\beta_{f}
a_{n}^{in}+\alpha_{f}^{*}b_{n}^{in\dagger},
\end{eqnarray}
where

\begin{eqnarray}
&&\beta_{f}=e^{-\pi\mu^{2}},
\\ && \nonumber
\alpha_{f}^{*}=-i\sqrt{\frac{2\pi}{\mu^{2}}}\frac{e^{-\pi\mu^{2}/2}}{\Gamma(i\mu^{2})},
\end{eqnarray}
with $\alpha_{f}$ and $\beta_{f}$ having the relation,

\begin{eqnarray}
|\alpha_{f}|^{2}+|\beta_{f}|^{2}=1. \label{fermrela}
\end{eqnarray}

We let $\alpha_{f}$ = $\cos{r_{f}}e^{i\phi}$ and $\beta_{f}$ =
$\sin{r_{f}}$, $r_{f}$ being a parameter with values between 0 and
${\pi}/{2}$ and related to the acceleration. Also, we can relate the
incoming states with the outgoing states as in the case of an
accelerating detector~\cite{nf},

\begin{eqnarray}
&&
\label{fermout}|0_{p}\rangle_{in}=\cos{r_{f}}e^{-i\phi}|0_{p}\rangle_{out}|0_{a}\rangle_{out}-
\sin{r_{f}}|1_{p}\rangle_{out}|1_{a}\rangle_{out},
\\ && \nonumber
|1_{p}\rangle_{in}=|1_{p}\rangle_{out}|0_{a}\rangle_{out}.
\end{eqnarray}

\subsection{Logarithmic Negativities}
The entanglement can be quantified by the logarithmic
negativity~\cite{ne_1,ne_2}. For a density operator $\rho_{A,B}$
corresponding to a bipartite system $A$ and $B$, we define the trace
norm $||\rho_{A,B}||$ $\equiv$ $tr|\rho_{A,B}|$ =
$tr\sqrt{\rho_{A,B}^{\dagger}\rho_{A,B}}$ and the negativity

\begin{eqnarray}
 N_{e}\equiv\frac{||\rho^{T_{A}}||-1}{2},
\end{eqnarray}
where $\rho^{T_{A}}$ is the partial transpose of $\rho_{A,B}$ with
respect to the party A. $N_{e}$ can be calculated from the absolute
value of the sum of the negative eigenvalues of $\rho^{T_{A}}$. Then
the logarithmic negativity of the bipartite system $A$ and $B$ is
defined by,

\begin{eqnarray}
LN(\rho_{A,B})\equiv\log_{2}||2N_{e}+1||.
\end{eqnarray}
For a product state, $LN(\rho_{A,B})$ = 0, and for entangled states,
$LN(\rho_{A,B})
>0$.

\section{Accelerating fermions}\label{sec4}

Initially, we have the incoming entangled state,
\begin{eqnarray}
\label{ferins}\Psi_{i}=\frac{1}{\sqrt{2}}\left[|0_{s,p}\rangle_{in}|0_{\omega,p}\rangle_{in}+
|1_{s,p}\rangle_{in}|1_{\omega,p}\rangle_{in}\right].
\end{eqnarray}
Then either one or both of the particles in $\omega$ and $s$ modes
are accelerated by the electric field, and the in states in
Eq.~(\ref{ferins}) are replaced by the out states as in
Eq.~(\ref{fermout}). If only the $\omega$ mode is accelerated, we
have

\begin{flushleft}
\begin{eqnarray}
\Psi_{f}  &=& \frac{1}{\sqrt{2}}\left\{|0_{s,p}\rangle_{out} \right.
 \otimes\left[\cos{r_{f}}e^{-i\phi}|0_{\omega,p}\rangle_{out}|0_{\omega,a}\rangle_{out}-
\sin{r_{f}}|1_{\omega,p}\rangle_{out}|1_{\omega,a}\rangle_{out}\right]
\\ && \nonumber \left.
+|1_{s,p}\rangle_{out}\otimes(|1_{\omega,p}\rangle_{out}|0_{\omega,a}\rangle_{out})\right\}.
\end{eqnarray}
\end{flushleft}
If both the $s$ and $\omega$ modes are accelerated with the same
$r_{f}$, we have
\begin{flushleft}
\begin{eqnarray}
\Psi_{f}
&=&\frac{1}{\sqrt{2}}\left\{\left[\cos{r_{f}}e^{-i\phi_{1}}|0_{s,p}\rangle_{out}|0_{s,a}\rangle_{out}
 \right.
-\sin{r_{f}}|1_{s,p}\rangle_{out}|1_{s,a}\rangle_{out}\right]
\\ && \nonumber
\otimes\left[\cos{r_{f}}e^{-i\phi_{1}}|0_{\omega,p}\rangle_{out}|0_{\omega,a}\rangle_{out}
-\sin{r_{f}}|1_{\omega,p}\rangle_{out}|1_{\omega,a}\rangle_{out}\right]
\\ && \nonumber \left.
+\left[(|1_{s,p}\rangle_{out}|0_{s,a}\rangle_{out})\otimes(|1_{\omega,p}\rangle_{out}|0_{\omega,a}\rangle_{out})\right]\right\}.
\end{eqnarray}
\end{flushleft}

The degradation of entanglement in the case of an accelerating
detector is due to the fact that some degrees of freedom have been
traced out. An accelerating detector 'sees' the space-time being
split into two causally disconnected regions, and it cannot access
information in one of them. We have verified explicitly that the
entanglement between the particles in $s$ mode and $\omega$ mode is
unchanged if there is no tracing out of any space-time region. On
the other hand, in the case of accelerating particles, the detector,
which is in an inertial frame, can access all degrees of freedom and
the orthogonality of the states is unchanged; therefore, the
entanglement of accelerating particles is unchanged.

However, more degrees of freedom are produced and we can calculate
the entanglements between different bipartite systems. In
Ref.~\cite{adami}, it was shown that entanglement is Lorentz
invariant. If one traces out the momentum, the entanglement
decreases, and the entanglement is transferred from the momentum to
the spin degrees of freedom. We will show that entanglement transfer
also occurs in accelerating fermions, from the particles to the
produced antiparticles.

\begin{figure}
\includegraphics{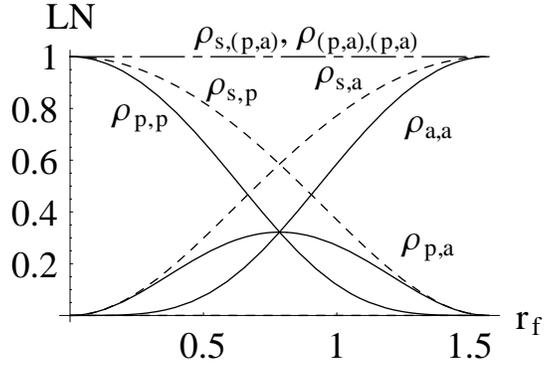}
\caption{\label{fig:bfacc} Logarithmic negativities of several
bipartite systems when one or both fermions are accelerated, the
magnitude of which is parameterized by $r_{f}$. In both cases, the
entanglement between the entire $s$ mode and $\omega$ mode is
unchanged (dot-dashed line). The solid lines show the results when
both particles are accelerated together, for three bipartite
systems: particles in $s$ mode and particles in $\omega$ mode
($\rho_{p,p}$), particles in $s$ mode and antiparticles in $\omega$
mode ($\rho_{p,a}$), and antiparticles in $s$ mode and antiparticles
in $\omega$ mode ($\rho_{a,a}$).  For comparison, the dashed lines
show the results when only the particle in the $\omega$ mode is
accelerated, in which case the two bipartite systems are particle in
$s$ mode and particles in $\omega$ ($\rho_{s,p}$), and particle in
$s$ mode and antiparticles in $\omega$ ($\rho_{s,a}$). }
\end{figure}

If only the particle in the $\omega$ mode is accelerated, we can
study the three bipartite systems: $A$ = the $s$ mode, $B$ = the
particles in the $w$ mode, the antiparticles in $w$ mode, or the
entire $w$ mode including both the particles and antiparticles. The
density matrices are called $\rho_{s,p}$, $\rho_{s,a}$, and
$\rho_{s,(p,a)}$ respectively. The entanglements are
\begin{eqnarray}
\left\{
  \begin{array}{ll}
    LN(\rho_{s,(p,a)})=1, &\ \\
    LN(\rho_{s,p})=\textrm{log}_{2}(1+\cos^{2}{r_{f}}), & \ \\
    LN(\rho_{s,a})=\textrm{log}_{2}(1+\sin^{2}{r_{f}}), &\
  \end{array}
\right.
\end{eqnarray}
which are plotted in Fig.~\ref{fig:bfacc}. It is obvious that the
entanglement of $\rho_{s,p}$ is transferred to $\rho_{s,a}$.

When both the $s$ and $\omega$ modes are accelerated with the same
$r_{f}$, we can calculate the entanglements between the five
bipartite systems: particles in $s$ mode and particles in $\omega$
mode ($\rho_{p,p}$), antiparticles in $s$ and antiparticles in
$\omega$ ($\rho_{a,a}$), antiparticles in $s$ and particles in
$\omega$ ($\rho_{a,p}$), particles in $s$ and antiparticles in
$\omega$ ($\rho_{p,a}$), and the entire $s$ and $\omega$ modes
($\rho_{(p,a),(p,a)}$). The logarithmic negativities are
\begin{eqnarray}
\left\{
  \begin{array}{ll}
    LN(\rho_{(p,a),(p,a)})=1,  \\
    LN(\rho_{p,p})=\textrm{log}_{2}\left[1+\cos^{4}{r_{f}}\right],  \\
    LN(\rho_{a,a})=\textrm{log}_{2}\left[1+\sin^{4}{r_{f}}\right],  \\
     LN(\rho_{p,a})=\textrm{log}_{2}\left[1+\cos^{2}{r_{f}}\sin^{2}{r_{f}}\right].
  \end{array}
\right.
\end{eqnarray}
By symmetry, $LN (\rho_{a,p}) = LN(\rho_{p,a})$. The results are
shown in Fig.~\ref{fig:bfacc}. The entanglement is transferred from
$\rho_{p,p}$ not only to $\rho_{p,a}$, but also to $\rho_{a,a}$. In
fact, when the acceleration of the particles tends to infinity, the
entanglement is completely transferred to between the antiparticles
$\rho_{a,a}$. Note that in both cases, the negativities of the
subsystems add up to the that of the total system, i.e.,
$N_{e}[\rho_{s,(p,a)}] = N_{e}[\rho_{s,p}]+N_{e}[\rho_{s,a}]$,
$N_{e}[\rho_{(p,a),(p,a)}] =
N_{e}[\rho_{p,p}]+N_{e}[\rho_{a,a}]+N_{e}[\rho_{p,a}]+N_{e}[\rho_{a,p}]$.

\section{Accelerating scalar particles}\label{sec3}

\subsection{Spectrum}

The relation between the in states and out states for scalar
particles is just the same as the relation between the Minkowski
states and Rindler states in ~\cite{tele,nb}. However, their spectra
are different. For both cases of accelerating particles with an
inertial observer and inertial particles with an accelerating
observer, the spectra are
\begin{eqnarray}
S_{\omega}=\bigskip_{in}\langle
0|a_{\omega}^{out\dagger}a_{\omega}^{out}|0\rangle_{in}=\sinh^{2}{r}.
\end{eqnarray}
The spectrum of accelerating particles with an inertial observer is
$S_{\omega}=\exp(-\pi m/a)$, where $a=E/m$ corresponds to the
acceleration of the particle in the classical limit. However, a
uniformly accelerating detector measures a spectrum
$S_{\omega}=1/[\exp({2\pi|\omega|}/{a})-1]$. In the classical limit,
the detectors observe the same particle trajectories; however, a
uniformly accelerating detector measures a different spectrum of
particles as that by an inertial detector on uniformly accelerating
particles.

\subsection{Entanglement}

For scalar particles, we have the same initially entangled state in
Eq.~(\ref{ferins}). If only the particle in $\omega$ mode is put in
a uniform electric field, the entangled state becomes

\begin{eqnarray}
 \label{entanout} \Psi_{f}&=&\frac{1}{\sqrt{2}} \left\{ |0_{p,s}\rangle_{out}\otimes
  \left[\frac{1}{\cosh{r}}\sum_{n=0}^{\infty}\tanh^{n}{r}|n_{p,\omega}\rangle_{out}|n_{a,\omega}\rangle_{out}\right]+
  \right.\\
 && \nonumber \left. |1_{p,s}\rangle_{out}\otimes
\left[\frac{1}{\cosh^{2}{r}}\sum^{\infty}_{n=0}\tanh^{n}{r}\sqrt{n+1}|(n+1)_{p,\omega}\rangle_{out}|n_{a,\omega}\rangle_{out}\right]
 \right\}.
\end{eqnarray}
If both particles in $\omega$ and $s$ modes are put in the electric
field with same $r$, we have,
\begin{eqnarray}
 &\Psi_{f}&=\frac{1}{\sqrt{2}} \left\{ \frac{1}{\cosh^{2}{r}}\left[\sum_{n=0}^{\infty}\tanh^{n}{r}|n_{p,s}\rangle_{out}|n_{a,s}\rangle_{out}\right]\otimes
  \left[\sum_{n=0}^{\infty}\tanh^{n}{r}|n_{p,\omega}\rangle_{out}|n_{a,\omega}\rangle_{out}\right]
  \right.\\
 && \nonumber \left. +\frac{1}{\cosh^{4}{r}}\left[\sum^{\infty}_{n=0}\tanh^{n}{r}\sqrt{n+1}|(n+1)_{p,s}\rangle_{out}|n_{a,s}\rangle_{out}\right]
\right.\\ \nonumber && \left.\otimes
\left[\sum^{\infty}_{n=0}\tanh^{n}{r}\sqrt{n+1}|(n+1)_{p,\omega}\rangle_{out}|n_{a,\omega}\rangle_{out}\right]
 \right\}.
\end{eqnarray}

More degrees of freedom have arisen due to pair production. Again,
the entanglement between the entire $s$ and $\omega$ modes remains
unchanged, such that $LN(\rho_{s,\omega}) =
LN(\rho_{s,(p,a)})=LN(\rho_{(p,a), (p,a)}) = 1$ for all $r$. This is
because the Bogoliubov transformation is linear, and the
orthogonality property of the states is unchanged.
\begin{figure}
\includegraphics{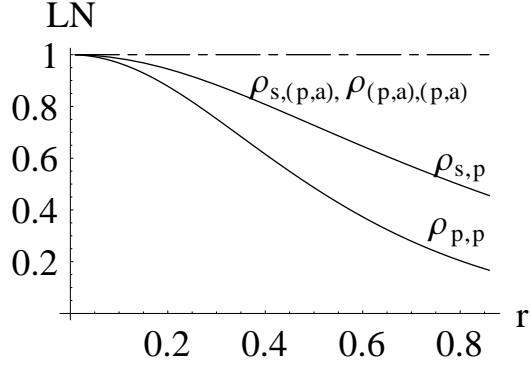}
\caption{\label{fig:enb_1} Same as Fig.~\ref{fig:bfacc}, but for
scalar particles.  Again, the entanglement of the entire $s$ mode
and $\omega$ mode, indicated by the dot-dashed line, remains
unchanged. Note that $LN(\rho_{s,a}) = LN(\rho_{p,a}) =
LN(\rho_{a,a}) = 0$ for all $r$. }
\end{figure}

We calculate the entanglements of $\rho_{s,p}$, $\rho_{s,a}$,
$\rho_{p,p}$, $\rho_{a,a}$ and $\rho_{p,a}$. To calculate the
density matrix, $\rho_{s,p}$~($\rho_{s,a}$), we trace over the
antiparticles~(particles). We take the partial transpose,
$\rho_{s,p}^{T}$, by interchanging the s mode's qubits to get an
infinite block-diagonal matrix. The ($n$, $n+1$) block matrix is,
\begin{eqnarray}
\frac{\tanh^{2n}{r}}{2\cosh^{2}{r}}\left(
  \begin{array}{cc}
    \frac{n}{\sinh^{2}{r}}  & \frac{\sqrt{n+1}}{\cosh{r}} \\
  \frac{\sqrt{n+1}}{\cosh{r}} & \tanh^{2}{r} \\
  \end{array}
\right).
\end{eqnarray}
Then we calculate the negative eigenvalues from each block matrix
and obtain the logarithmic negativity of $\rho_{s,p}$,

\begin{eqnarray}
LN(\rho_{s,p})&=&\textrm{log}_{2}\left[\frac{1}{2\cosh^{2}{r}}+
\sum_{n=0}^{\infty}\frac{\tanh^{2n}{r}}{2\cosh^{2}{r}}\sqrt{\left(\frac{n}{\sinh^{2}{r}}+
\tanh^{2}{r}\right)^{2}+\frac{4}{\cosh^{2}{r}}}\right].
\end{eqnarray}
We also calculate the $LN(\rho_{p,p})$ numerically. The results are
shown in Fig.~\ref{fig:enb_1}.

In contrast to fermions, there is no entanglement transfer to the
antiparticles for scalar particles, and $LN(\rho_{s,a}) =
LN(\rho_{p,a}) = LN(\rho_{a,a}) = 0$ for all $r$, even though the
entanglement between the particles in the $s$ and $\omega$ modes
decreases as $r$ increases.

\subsection{Entanglements if the Number of Produced Pairs is Restricted}

However, if we constrain the number of pairs produced,
$LN(\rho_{s,a})$, $LN(\rho_{p,a})$ and $LN(\rho_{a,a})$ are all
nonzero. If only $M$ pairs can be produced in a mode,

\begin{eqnarray}
&&
|0\rangle_{in}=\frac{N_{1}}{\cosh{r}}\sum_{n=0}^{M}\tanh^{n}{r}|n_{p}\rangle_{out}|n_{a}\rangle_{out},
\\ && \nonumber
|1_{p}\rangle_{in}
=\frac{N_{2}}{\cosh^{2}{r}}\sum^{M-1}_{n=0}\tanh^{n}{r}\sqrt{n+1}|(n+1)_{p}\rangle_{out}|n_{a}\rangle_{out},
\end{eqnarray}
where $N_{1}$ and $N_{2}$ are normalization factors,

\begin{eqnarray}
&& N_{1}=\left({1-\tanh^{2M+2}{r}}\right)^{-1/2},
\\ && \nonumber
N_{2}=\left[{{1-(M+1)\tanh^{2M}{r}+M\tanh^{2M+2}{r}}}\right]^{-1/2}.
\end{eqnarray}

\begin{figure}
\includegraphics{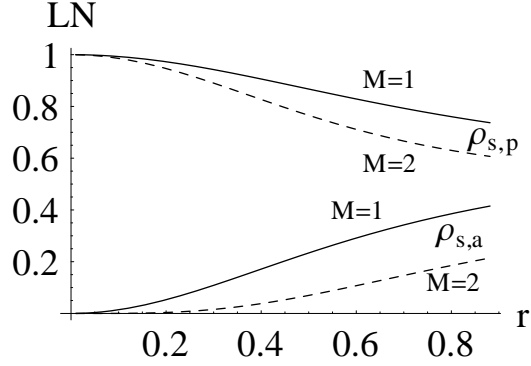}
\caption{\label{fig:encp_1} Same as Fig.~\ref{fig:enb_1}, but for
$M$ = 1 (solid lines) and $M$ = 2 (dashed lines), and $\rho_{s,a}$
is shown as well.}
\end{figure}

We take the partial transpose of $\rho_{s,p}$ which has $M$ diagonal
block matrices and the $n$th block is,
\begin{eqnarray}
\frac{\tanh^{2n-4}{r}}{2\cosh^{2}{r}}\left(\begin{array}{cc}
   \frac{N_{2}^{2}}{\cosh^{2}{r}}(n-1) & \frac{N_{2}N_{1}\tanh^{2}{r}}{\cosh{r}}\sqrt{n} \\
   \frac{N_{2}N_{1}\tanh^{2}{r}}{\cosh{r}}\sqrt{n} &
   N_{1}^{2}\tanh^{4}r
 \end{array}
\right).
\end{eqnarray}
We then sum up the negative eigenvalues of the $n$th blocks to
calculate the logarithmic negativity

\begin{eqnarray}
 LN(\rho_{s,p})&=& \textrm{log}_{2}\left\{1-
\sum_{n=1}^{M}\frac{\tanh^{2n-4}{r}}{2\cosh^{2}{r}}\left[\frac{(n-1)N_{2}^{2}}{\cosh^{2}{r}}+
{N_{1}^{2}\tanh^{4}{r}}- \right. \right.\\ && \nonumber \left.\left.
\sqrt{\left[\frac{(n-1)N_{2}^{2}}{\cosh^{2}{r}}+
{N_{1}^{2}\tanh^{4}{r}}\right]^{2}+
\frac{4N_{1}^{2}N_{2}^{2}\tanh^{4}{r}}{\cosh^{2}{r}}}\right]\right\}.
\end{eqnarray}
The partial transpose of $\rho_{s,a}$ also has a block diagonal
structure and only the last block
\begin{eqnarray}
\frac{\tanh^{2n-2}{r}}{2\cosh^{2}{r}}\left(
  \begin{array}{cc}
    N_{1}^{2} & \frac{N_{1}N_{2}\tanh{r}}{\cosh{r}}\sqrt{n} \\
    \frac{N_{1}N_{2}\tanh{r}}{\cosh{r}}\sqrt{n} & 0 \\
  \end{array}
\right)
\end{eqnarray}
contributes to the negative eigenvalue. Then the logarithmic
negativity of $\rho_{s,a}$ is
\begin{eqnarray}
LN(\rho_{s,a}) &=& \textrm{log}_{2}\left\{1-
\frac{N_{1}^{2}\tanh^{2M-2}{r}}{2\cosh^{2}{r}}\left[1-
\sqrt{1+\frac{4N_{2}^{2}M\tanh^{2}{r}}{N_{1}^{2}\cosh^{2}{r}}}\right]\right\}.
\end{eqnarray}
We show the results $M$ = 1 and 2 in Fig.~\ref{fig:encp_1}. In both
cases, $LN(\rho_{s,(p,a)})$ is equal to 1.

\begin{figure}
\includegraphics{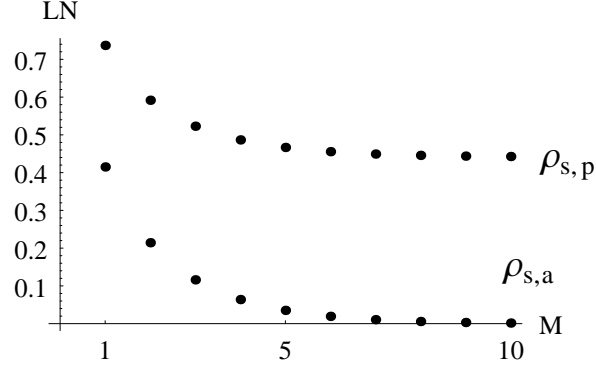}
\caption{\label{fig:nop_tp} Logarithmic negativity of $\rho_{s,p}$
and $\rho_{s,a}$ at infinite acceleration (\textit{r} = 0.88) as a
function of the number of produced pairs $M$.}
\end{figure}

When $M$ is finite, the entanglement of $\rho_{s,a}$ is not zero and
increases with $r$ while that of $\rho_{s,p}$ decreases with $r$.
However, the entanglements of both $\rho_{s,p}$ and $\rho_{s,a}$ are
reduced if more particles are produced~($M$ increases). The
dependence of $LN(\rho_{s,p})$ and $LN(\rho_{s,a})$ on $M$ at
infinite acceleration are shown in Fig.~\ref{fig:nop_tp}. Although
the entanglement of $\rho_{s,p}$ decreases more for greater $M$, the
entanglement of $\rho_{s,a}$ also decreases with a similar trend and
goes to zero when $M\rightarrow \infty$. Therefore, there is no
transfer of entanglement to the antiparticles for scalar particles
when the number of produced pairs is not restricted.

We next calculate the case when both particles in $\omega$ and $s$
modes are put in the uniform electric field such that they are both
accelerated and have the same $r$. In the out basis, the state in
Eq.~(\ref{ferins}) becomes,
\begin{eqnarray}
 &&\Psi_{f}=\frac{1}{\sqrt{2}} \left\{ \frac{N_{1}^{2}}{\cosh^{2}{r}}\left[\sum_{n=0}^{M}\tanh^{n}{r}|n_{p,s}\rangle_{out}|n_{a,s}\rangle_{out}\right]\otimes
  \left[\sum_{n=0}^{M}\tanh^{n}{r}|n_{p,\omega}\rangle_{out}|n_{a,\omega}\rangle_{out}\right]+
  \right.\\
 && \nonumber \left. \frac{N_{2}^{2}}{\cosh^{4}{r}}\left[\sum^{M}_{n=0}\tanh^{n}{r}\sqrt{n+1}|(n+1)_{p,s}\rangle_{out}|n_{a,s}\rangle_{out}\right]\otimes
\left[\sum^{M}_{n=0}\tanh^{n}{r}\sqrt{n+1}|(n+1)_{p,\omega}\rangle_{out}|n_{a,\omega}\rangle_{out}\right]
 \right\}.
\end{eqnarray}

\begin{figure}
\includegraphics{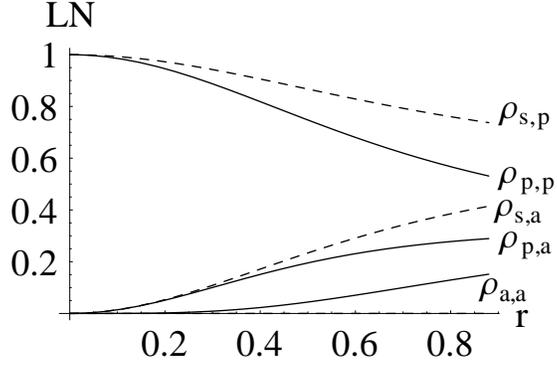}
\caption{\label{fig:bsacc_1} Logarithmic negativities of different
bipartite systems when the particles in both $\omega$ and $s$ modes
are accelerated, indicated by $\rho_{p,p}$~(between the particles
and particles), $\rho_{p,a}$~(between particles in $s$ and
antiparticles in $\omega$ modes), and $\rho_{a,a}$~(between
antiparticles in $s$ and $\omega$ modes), for $M=1$. For comparison,
the previous results~(shown in Fig.~\ref{fig:encp_1}) when only the
particle in $\omega$ is accelerated are plotted as dashed lines.}
\end{figure}

\begin{figure}
\includegraphics{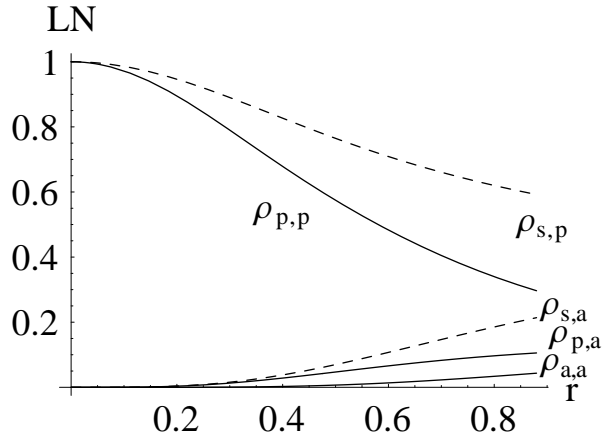}
\caption{\label{fig:bsacc_2} Same as Fig.~\ref{fig:bsacc_1}, but for
$M=2$.}
\end{figure}

Now, we can consider even more bipartite systems between the two
modes. We calculate the entanglements between the particles or
antiparticles in the $s$ mode and the particles or antiparticles in
the $\omega$ mode,~i.e.,  of $\rho_{p,p}$, $\rho_{a,a}$,
$\rho_{p,a}$ and $\rho_{a,p}$ for the cases $M=1$ and $2$. As
expected from symmetry, $LN(\rho_{p,a})$ is equal to
$LN(\rho_{a,p})$. The results are shown in Figs.~\ref{fig:bsacc_1}
and \ref{fig:bsacc_2} for $M=1$ and $2$ respectively. As in the case
when only one of the particles is accelerated, the entanglement
between the particles in $s$ and $\omega$ modes is degraded when $M$
increases. It seems that entanglement transfer also occurs for
scalar particles for finite $M$. However, in this case we note that
the negativities of the $\rho_{a,a}$, $\rho_{p,a}$,  $\rho_{a,p}$,
and $\rho_{p,p}$ do not sum to a constant, and we cannot identify a
causal relation between the decrease in entanglement between the
particles and the increase in entanglement between the
antiparticles. A more rigorous definition of entanglement transfer
is needed before we can discuss the issue for scalar particles
further.

\section{Conclusion}\label{sec5}
We have studied how the entanglement of a pair of particles is
affected when one or both of the pair is uniformly accelerated, as
measured by an inertial detector, and compared the results with that
of inertial particles observed by a uniformly accelerating detector.
While there is a degradation of entanglement in the latter case due
to the splitting of the space-time, the entanglement in the former
case is unchanged by the acceleration when all degrees of freedom
are considered. Furthermore, the spectrum of the uniformly
accelerating particles is different from that seen by uniformly
accelerating detectors.

When one of the particles - the one in $\omega$ mode - is uniformly
accelerated, the entanglement is transferred to the produced
antiparticles for fermions, while there is no such entanglement
transfer for scalar particles. For scalar particles, when the number
of produced pairs is restricted, the entanglement of $\rho_{s,a}$
increases with $r$. However, at any $r$, $LN(\rho_{s,a})$ decreases
as $M$ increases and goes to zero when $M\rightarrow \infty$.

When both particles in $s$ and $\omega$ modes are uniformly
accelerated~(with the same $r$ or $r_{f}$), we have even more
bipartite systems. For fermions, $\rho_{a,a}$ takes up all the
entanglement at large acceleration. For scalar particles with
restricted number of produced pairs, the entanglements of
$\rho_{p,a}$ and $\rho_{a,a}$ increase with $r$. However, if there
is no restriction of the number of produced pairs, no entanglement
transfer to the antiparticles is observed for scalar particles.

Our results raise the possibility that when an entangled pair falls
into a black hole, their entanglement may be partially transferred
to the produced particles, which should not be ignored in
considering the black hole information paradox. Studying quantum
entanglement in curved space-time may therefore give us insights on
the relation between quantum mechanics and general relativity.

\appendix
\section{The bogoliubov coefficients in the scalar case} \label{appa}

We assume the form of solution of Eq.~(\ref{okg}) as
\begin{eqnarray}
\phi_{\omega}(t,x)=Ce^{i\omega t}\chi_{\omega}(x),
\end{eqnarray}
where $C$ is a normalization constant, and we obtain from
Eq.~(\ref{okg})

\begin{eqnarray}
\left[\frac{\partial^{2}}{\partial
x^{2}}+E^{2}(x-\omega/E)^{2}\right]\chi_{\omega}(x)=m^{2}\chi_{\omega}(x).
\label{simkg}
\end{eqnarray}
The solutions of Eq.~(\ref{simkg}) can be found in ~\cite{hand}, and
they are parabolic cylinder functions,

\begin{eqnarray}
&& D_{i\mu^{2}-\frac{1}{2}}[\pm(1+i)\sqrt{E}(x-\omega)/E],
\\ \nonumber &&
D_{-i\mu^{2}-\frac{1}{2}}[\pm(1-i)\sqrt{E}(x-\omega)/E].
\end{eqnarray}
We can use the saddle point method to classify the solutions in the
in/out basis ~\cite{kgwp_0,greiner} and have the in-basis functions,
\begin{eqnarray}
&&\phi^{in}_{\omega,p}(x,t)
=\frac{e^{-3\pi\mu^{2}/4}}{(2E)^{1/4}}e^{i\omega
t}D_{i\mu^{2}-1/2}[e^{-3i\pi/4}\sqrt{2E}(x-\omega/E)],
\\ &&
\phi^{in}_{\omega,a}(x,t)=\phi^{in}_{-\omega,p}(-x,t),
\end{eqnarray}
where $\mu^{2}=m^{2}/2E$. The subscripts $p$ and $a$ stand for
particles and antiparticles respectively. We also obtain the
out-basis solutions,

\begin{eqnarray}
&& \phi^{out}_{\omega,p}(x,t)=\phi^{in*}_{\omega,p}(x,-t),
\\ &&
\phi^{out}_{\omega,a}(x,t)=\phi^{in*}_{-\omega,p}(-x,-t).
\end{eqnarray}
The solutions have been normalized by the Klein-Gordon scalar
product,

\begin{eqnarray}
&&\int dx
\phi_{\omega^{'},p}^{in*}(x,t)i\overleftrightarrow{D_{t}}\phi_{\omega,p}^{in}(x,t)=+\delta(\omega-\omega^{'}),
\\ \nonumber &&
\int
dx\phi_{\omega^{'},a}^{in*}(x,t)i\overleftrightarrow{D_{t}}\phi_{\omega,a}^{in}(x,t)=-\delta(\omega-\omega^{'}).
\end{eqnarray}
As there are two different complete bases, we can quantize the field
in two ways,

\begin{eqnarray}
&&
\phi=\sum_{\omega}(a_{\omega}^{in}\phi_{\omega,p}^{in}+b_{\omega}^{in\dagger}\phi_{\omega,a}^{in*}),
\end{eqnarray}
or
\begin{eqnarray}
\phi=\sum_{\omega}(a_{\omega}^{out}\phi_{\omega,p}^{out}+b_{\omega}^{out\dagger}\phi_{\omega,a}^{out*}).
\end{eqnarray}
From~\cite{hand}, we can get a relation between the out-basis and
in-basis solutions,
\begin{eqnarray}
\label{relationink}&&D_{-\frac{1}{2}-i\mu^{2}}[-(1+i)\sqrt{E}(x-\omega/E)]
\\ && \nonumber=
e^{-\pi\mu^{2}+\frac{i\pi}{2}}D_{-\frac{1}{2}-i\mu^{2}}[(1+i)\sqrt{E}(x-\omega/E)]+\frac{\sqrt{2\pi}e^{-i\pi/4}e^{-\pi\mu^{2}/2}}{\Gamma(\frac{1}{2}+i\mu^{2})}
D_{-\frac{1}{2}+i\mu^{2}}[-(1-i)\sqrt{E}(x-\omega/E)].
\end{eqnarray}
We write Eq.~(\ref{relationink}) in the form of in-basis and
out-basis solutions explicitly,
\begin{eqnarray}
\phi_{\omega,p}^{out}=\beta\phi_{\omega,a}^{in*}+\alpha\phi_{\omega,p}^{in}.
\end{eqnarray}
Since $\beta=e^{i\phi_{2}}\sinh{r}$, $r= \sinh^{-1}[\exp(-\pi
m/2a)]$.

\section{Reducing the Dirac equations to two Klein-Gordon equations}\label{appb}

From Eq.~(\ref{diraceqt}), we let
\begin{eqnarray}
\psi=(\gamma^{\nu}(p_{\mu}-A_{\mu})+m)\phi \label{diraclet}
\end{eqnarray}
to obtain
\begin{eqnarray}
\label{diracsim} [(p-A)^{2}-m^{2}-{i}\sigma^{\mu
\nu}(p_{\mu}-A_{\mu})(p_{\nu}-A_{\nu})]\phi=0,
\end{eqnarray}
where $\sigma^{\mu \nu}={i[\gamma^{\mu},\gamma^{\nu}]}/{2}$. We
choose the gauge to be $A_{0}=0$, $A_{3}=-Et$. We then substitute
the potential in Eq.~(\ref{diracsim}) to get
\begin{eqnarray}
\left[-\frac{\partial^{2}}{\partial
t^{2}}-\left(i\frac{\partial}{\partial
z}+Et\right)^{2}-m^{2}+iE\alpha_{3}\right]\phi=0,
\end{eqnarray}
where
\begin{eqnarray}
\alpha_{3}=\left(
             \begin{array}{cc}
               0 & \sigma_{3} \\
               \sigma_{3} & 0 \\
             \end{array}
           \right),
\end{eqnarray}
and $\sigma_{3}$ is the Pauli matrix. We then assume that the
solution has the form
\begin{eqnarray}
\phi=e^{ik z}n(t),
\end{eqnarray}
where
\begin{eqnarray}
n(t)=\sum^{4}_{\lambda=1}f_{\lambda}(t)u_{\lambda},
\end{eqnarray}
with the spinors,
\begin{eqnarray}
&& u_{1}=\frac{1}{\sqrt{2}}\left(
                   \begin{array}{c}
                     1 \\
                     0 \\
                     1 \\
                     0 \\
                   \end{array}
                 \right),u_{2}=\frac{1}{\sqrt{2}}\left(
                                                   \begin{array}{c}
                                                     0 \\
                                                     1 \\
                                                     0 \\
                                                     -1 \\
                                                   \end{array}
                                                 \right),
                                                 \\ && \nonumber
                                                 u_{3}=\frac{1}{\sqrt{2}}\left(
                                                                           \begin{array}{c}
                                                                             1 \\
                                                                             0 \\
                                                                             -1 \\
                                                                             0 \\
                                                                           \end{array}
                                                                         \right),
                                                                         u_{4}=\frac{1}{\sqrt{2}}\left(
                                                                                                   \begin{array}{c}
                                                                                                     0 \\
                                                                                                     1 \\
                                                                                                     0 \\
                                                                                                     1 \\
                                                                                                   \end{array}
                                                                                                 \right).
\end{eqnarray}

With the relations,
\begin{eqnarray}
\alpha_{3}u_{\lambda}=\eta u_{\lambda}, \left\{
                                        \begin{array}{ll}
                                           \eta=1, & \hbox{for } \lambda =1, 2, \\
                                         \eta =-1, & \hbox{for }\lambda
=3,4,
                                        \end{array}
                                      \right.
\end{eqnarray}
we can get two Klein-Gordon equations,
\begin{eqnarray}
\left[\frac{\partial^{2}}{\partial
t^{2}}+E^{2}\left(t-\frac{k}{E}\right)^{2}+m^{2}-iE\eta
\right]f_{\lambda}(t)=0.
\end{eqnarray}
The solutions are parabolic cylinder functions.

We can then classify the in/out solutions~\cite{kgwp_1, Aniki}.
$f_{3}$ and $f_{4}$ are dependent on $f_{1}$ and $f_{2}$, and so we
just consider the cases of $\lambda=1, 2$ in the following.
Neglecting the normaliazation factors, we write down the in/out
solutions,

\begin{eqnarray}
\label{ds1}\begin{array}{ll}
     &  \phi_{p}^{in}=e^{ikz}D_{i\mu^{2}}[-(1-i)\sqrt{E}(t-k/E)],\\
     &  \phi_{a}^{in}=e^{ikz}D_{-i\mu^{2}-1}[-(1+i)\sqrt{E}(t-k/E)],\\
     & \phi_{p}^{out}=e^{ikz}D_{-i\mu^{2}-1}[(1+i)\sqrt{E}(t-k/E)], \\
     & \phi_{a}^{out}=e^{ikz}D_{i\mu^{2}}[(1-i)\sqrt{E}(t-k/E)],
  \end{array}
\end{eqnarray}
where $\mu^{2}=m^{2}/2E$. We substitute the solutions in
Eq.~(\ref{ds1}) into Eq.~(\ref{diraclet}), to obtain the solutions
in the Dirac equation, i.e., Eq.~(\ref{diraceqt}). We show the
calculation of $\phi_{p}^{in}$ with $\lambda =1$ as follows.

\begin{eqnarray}
\psi_{p,1}^{in}&=&\left[\gamma^{\mu}(p_{\mu}-A_{\mu})+m\right]u_{1}e^{ikz}D_{i\mu^{2}}[-(1-i)\sqrt{E}(t-k/E)]
\\ \nonumber &=&
e^{ikz}\left[mu_{1}D_{i\mu^{2}}[-(1-i)\sqrt{E}(t-{k}/{E})]-(1-i)\mu^{2}\sqrt{E}u_{1}^{'}
D_{i\mu^{2}-1}[-(1-i)\sqrt{E}(t-{k}/{E})]\right],
\end{eqnarray}
where $u_{1}^{'}=\gamma^{0}u_{1}=\gamma^{3}u_{1}$.  We normalize it
and calculate $\psi_{a}^{in}$, $\psi_{p}^{out}$ and
$\psi_{a}^{out}$. We can relate the solutions using two mathematical
relations

\begin{eqnarray}
&& D_{i\mu^{2}-1}[(1-i)x]=-e^{-{\pi
\mu^{2}}}D_{i\mu^{2}-1}[-(1-i)x]+\frac{\sqrt{2\pi}e^{-\pi
\mu^{2}/2}}{\Gamma(1-{i\mu}^{2})}D_{-i\mu^{2}}[-(1+i)x],
\\ \nonumber &&
 D_{i\mu^{2}}[(1-i)x]=e^{-{\pi
\mu^{2}}}D_{i\mu^{2}}[-(1-i)x]+\frac{i\sqrt{2\pi}e^{-\pi
\mu^{2}/2}}{\Gamma(-{i\mu}^{2}/{2})}D_{-1-i\mu^{2}}[-(1+i)x],
\end{eqnarray}
and so we can obtain the Bogoliubov coefficients,
\begin{eqnarray}
&&\beta_{f}=e^{-\pi\mu^{2}},
\\ && \nonumber
\alpha_{f}^{*}=-i\sqrt{\frac{2\pi}{\mu^{2}}}\frac{e^{-\pi\mu^{2}/2}}{\Gamma(i\mu^{2})}.
\end{eqnarray}
Since $\beta_{f}= \sin{r_{f}}$, $r_{f}=\arcsin{[\exp(-\pi m/2a)]}$.


\begin{thebibliography}{99}
\small

        \bibitem{qutele} C. H. Bennett \textit{et al}., Phys. Rev. Lett. \textbf{70}, 1895 (1993).

        \bibitem{qc}S. F. Huelga, M. B. Plenio, and J. A. Vaccaro, Phys. Rev. A \textbf{65}, 042316 (2002).

        \bibitem{qs} J. L. Dodd, M. A. Nielsen, M. J. Bremner, and R. T. Thew, Phys. Rev. A \textbf{65}, 040301(R) (2002).


        \bibitem{adami} R. M. Gingrich and C. Adami, Phys. Rev. Lett. \textbf{89}, 270402 (2002).
        \bibitem{alsing} P. M. Alsing and G. J. Milburn, Quantum Inf. Comput. \textbf{2}, 487 (2002); eprint quant-ph/0203051.

        \bibitem{tele} P. M. Alsing and G. J. Milburn, Phys. Rev. Lett. \textbf{91}, 180404 (2003).
        \bibitem{nb}I. Fuentes-Schuller and R. B. Mann, Phys. Rev. Lett. \textbf{95}, 120404 (2005).
        \bibitem{nf}P. M. Alsing, I. Fuentes-Schuller, R. B. Mann, and T. E. Tessier,
        quant-ph/0603269v2.
        \bibitem{unruh}W. G. Unruh, Phys. Rev. D \textbf{14}, 870 (1976).
        \bibitem{zhang} Y. Ling \textit{et al}., J. Phys. A: Math. Theor. \textbf{40}, 9025
        (2007).



        \bibitem{sd} S. Parker \textit{et al}., Phys. Rev. A \textbf{61}, 032305 (2000).
        \bibitem{kgwp_0}R. Brout, S. Massar, R. Parentani, and Ph. Spindel, Phys. Rep. \textbf{260}, 329 (1995).
        \bibitem{lne} G. Vidal and R. F. Werner, Phys. Rev. A \textbf{65}, 032314 (2002).




        \bibitem{accwp} G. Vandegrift, Am. J. Phys. \textbf{68}, 576 (2000).




        \bibitem{fs} F. Sauter, Z. Phys. \textbf{69}, 742 (1931).

        \bibitem{wh}  W. Heisenberg and H. Euler, Z. Phys. \textbf{98}, 714 (1936).

        \bibitem{jsch} J. Schwinger, Phys. Rev. \textbf{82}, 664 (1951).

        \bibitem{pptg_1} T. Padmanabhan, Pramana-Journal of Physics \textbf{37}, 179 (1991).
        \bibitem{pptg_2}K. Srinivasan and T. Padmanabhan, Phys. Rev. D \textbf{60}, 024007 (1999).
        \bibitem{ppr}Cl. Gabriel and Ph. Spindel, Ann. Phys. (N.Y.) \textbf{284}, 263 (2000).

        \bibitem{ppf_1}R.-C. Wang and C.-Y. Wong, Phys. Rev. D \textbf{38}, 348 (1988).
        \bibitem{ppf_2}C. Martin and D. Vautherin, Phys. Rev. D \textbf{38}, 3593 (1988).
        \bibitem{ppf_3}C. Martin and D. Vautherin, Phys. Rev. D \textbf{40}, 1667 (1989).


        \bibitem{kgwp_1} R. Brout, S. Massar, R. Parentani, S. Popescu, and Ph. Spindel, Phys. Rev. D \textbf{52}, 1119 (1995).

        \bibitem{klein}O. Klein, Z. Phys. \textbf{53}, 157 (1929).

        \bibitem{greiner}W. Greiner, B. M\"{u}ller, and J. Rafelski, \textit{Quantum Electrodynamics of Strong Fields} (Springer, Berlin, 1985).

        \bibitem{niki} A. Nikishov, Journal of Experimental and Theoretical Physics \textbf{96}, 180 (2003).

        \bibitem{ne_1}M. B. Plenio, Phys. Rev. Lett. \textbf{95}, 090503 (2005).
        \bibitem{ne_2}M. B. Plenio and S. Virmani, quant-ph/0504163v3.

        \bibitem{hand}M. Abramowitz and I. A. Stegun, \textit{Handbook of Mathematical Functions} (Dover, New York, 1964).
        \bibitem{Aniki} A. I. Nikishov, Journal of Russian Laser
        Research \textbf{6}, 1573 (1985).


\end{thebibliography}
\end{document}